\documentclass[runningheads]{llncs}
\authorrunning{Ahmadi et. al}
\usepackage{amsmath,amsfonts,amssymb,mathtools}
\usepackage[ruled,lined,linesnumbered,commentsnumbered,noend]{algorithm2e}
\usepackage{float}
\usepackage{setspace}
\newcommand{\norm}[1]{\left\lVert#1\right\rVert}
\DeclareMathOperator*{\argmin}{arg\,min}
\newcommand\given[1][]{\:#1\vert\:}
\DeclareMathOperator{\E}{\mathbb{E}}
\usepackage{graphicx}
\usepackage{thmtools}
\usepackage{thm-restate}
\usepackage{hyperref}

\declaretheorem[name=Lemma]{lem}
\declaretheorem[name=Theorem]{thm}

\newenvironment{proof-sketch}{\noindent{\bf Sketch of Proof}\hspace*{1em}}{\qed\bigskip}
\allowdisplaybreaks
\title{Min-Max Correlation Clustering via MultiCut\protect\footnote{
A preliminary version of this work appeared in IPCO 2019.~\cite{prelim-version}.\\ 
The first and third authors are supported by NSF grant CNS 156019. Part of the research was done when the fourth author was visiting the Simons Institute of Theory of Computing and the author is supported by NSF CAREER 1652303, NSF CCF 1464310 and a Google faculty award.
The last author is supported by the ISF (grant No. 1336/16).}}

\newif\ifjournal
\journaltrue

\begin{document}
%

\author{Saba Ahmadi\inst{1}\and Sainyam Galhotra\inst{2}\and
Samir Khuller\inst{3} \and
Barna Saha\inst{2}\and Roy Schwartz\inst{4}}
%
%
\institute{University of Maryland, College Park\\
\email{\{saba\}@cs.umd.edu}\\
\and College of Information \& Computer Science\\
University of Massachussetts Amherst\\
\email{sainyam,barna@cs.umass.edu}\\
\and Northwestern University\\
\email{\{samir.khuller\}@northwestern.edu}\\
\and Technion -- Israel Institute of Technology\\
\email{\{schwartz\}@cs.technion.ac.il}}

\maketitle 
\begin{abstract}
Correlation clustering is a fundamental combinatorial optimization problem arising in many contexts and 
applications that has been the subject of dozens of papers
in the literature.
In this problem we are given a general weighted graph where each edge is labeled positive or negative. The goal is to 
obtain a partitioning (clustering) of the vertices that minimizes disagreements -- weight of negative edges trapped inside a cluster plus positive edges between different clusters.
Most of the papers on this topic mainly focus on minimizing
total disagreement, a global objective for this problem.

In this paper we study a cluster-wise objective function that asks to minimize the maximum number of disagreements of each cluster, which we call min-max correlation clustering. The min-max objective is a natural objective that respects the quality of every cluster. In this paper, we provide the first nontrivial approximation algorithm for this problem achieving an $\mathcal{O}(\sqrt{\log n\cdot\max\{\log(|E^-|),\log(k)\}})$ approximation for general weighted graphs, where $|E^-|$ denotes the number of negative edges and $k$ is the number of clusters in the optimum solution. To do so, we also obtain a corresponding result for multicut where we wish to find a multicut solution while trying to minimize the total weight of cut edges on every component. The results are then further improved to obtain (i) $\mathcal{O}(r^2)$-approximation for min-max correlation clustering and min-max multicut for graphs that exclude $K_{r,r}$ minors
(ii) a 14-approximation for the min-max correlation clustering on complete graphs. 
\end{abstract}
\vspace{-0.9cm}
\section{Introduction}
\vspace{-0.3cm}
Correlation clustering is a fundamental optimization problem introduced by Bansal, Blum and Chawla \cite{bansal2004correlation}. In this problem, we are given a general weighted graph where each edge is labeled positive or negative. The goal is to obtain a partitioning of the vertices into an arbitrary number of clusters that agrees with the edge labels as much as possible. That is, a clustering that minimizes disagreements, which is the weight of positive edges between the clusters plus the weight of negative edges inside the clusters. In addition, correlation clustering captures some fundamental graph cut problems including min s-t cut, multiway cut and multicut.
Correlation clustering has been studied extensively for more than a decade~\cite{ailon2012improved,ailon2008aggregating,charikar2003clustering,chawla2015near,demaine2006correlation}. Most of the papers have focused on a global min-sum objective function, i.e. minimizing total number of disagreements or maximizing the total number of agreements.


In recent work, Puleo and Milenkovic \cite{puleo2015correlation} introduced a local vertex-wise min-max objective for correlation clustering which bounds the maximum number of disagreements on each node. This problem arises in many community detection applications in machine learning, social sciences, recommender systems and bioinformatics \cite{cheng2000biclustering,symeonidis2006nearest,kriegel2009clustering}. This objective function makes sure each individual has a minimum quality within the clusters.
They showed this problem is NP-hard even on un-weighted complete graphs, and developed an $O(1)$ approximation
algorithm for unweighted complete graphs. Charikar et al. \cite{charikar2017local} improved upon the work by Puleo et al. \cite{puleo2015correlation} for complete graphs by giving a 7 approximation. For general weighted graphs, their approximation bound is $O(\sqrt{n})$ where $n$ is the number of vertices. Both these
algorithms rely on LP rounding, based on a standard linear program relaxation for the problem. In contrast, for the global minimization objective an $O(\log{n})$-approximation can be obtained \cite{demaine2006correlation}. Therefore, the local objective for correlation clustering seems significantly harder to approximate than the global objective.

In this work, we propose a new local cluster-wise min-max objective for correlation clustering -- minimizing the maximum number of disagreements of each cluster. 
This captures the case when we wish to 
create communities that are harmonious, as global min sum objectives
could create an imbalanced community structure. This new local objective guarantees fairness to communities instead of individuals. To name a few applications for this new objective, consider a task of instance segmentation in an image which can be modeled using correlation clustering ~\cite{kirillov2017instancecut,NIPS2011_4406}. A cluster-wise min-max objective makes sure each detected instance has a minimum quality. Another example is in detecting communities in social networks, this objective makes sure there are no communities with lower quality compared to the other communities. No hardness results are known for the cluster-wise min-max objective.
A similar objective was proposed for the multiway cut problem by Svitkina and Tardos~\cite{svitkina2004min}. In the min-max multiway cut problem, given a graph $G$ and $k$ terminals the goal is to get a partitioning of $G$ of size $k$ that separates all terminals and the maximum weight of cut edges on each part is minimized. Svitkina and Tardos~\cite{svitkina2004min} showed an $\mathcal{O}(\log^3 n)$ approximation algorithm for min-max multiway cut on general graphs (this bound immediately improves to $\mathcal{O}(\log^2 n)$ by using better bisection algorithms).
Bansal et al. \cite{bansal2014min} studied a graph partitioning problem called min-max $k$-partitioning from a similar perspective. In this problem, given a graph $G=(V,E)$ and $k\geq 2$ the goal is to partition the vertices into $k$ roughly equal parts $S_1,\cdots,S_k$ while minimizing $\max_i\delta(S_i)$. They showed an $\mathcal{O}(\sqrt{\log n \log k})$ approximation algorithm for this problem. They also improved the approximation ratio given by Svitkina et al. \cite{svitkina2004min} for min-max multiway cut to $\mathcal{O}(\sqrt{\log n \log k})$.
Bansal et al's seminal work \cite{bansal2014min} uses the concept of orthogonal
separators introduced by Chlamtac et al.~\cite{chlamtac2006play} to achieve
their result.

\vspace{-0.4cm}
\section{Results \& High Level Ideas}
\vspace{-0.3cm}
In this paper, we give an approximation algorithm for the problem of min-max correlation clustering.
\vspace{-0.1cm}
\begin{definition}
(Min-max Correlation Clustering) Let $G=(V,E)$ be an edge-weighted graph such that each edge is labeled positive or negative. The min-max correlation clustering problem asks for a
partioning of the nodes (a clustering) where the maximum disagreement of a cluster is minimized. Disagreement of a cluster $C$ is the weight of negative edges with both endpoints inside $C$ plus the weight of positive edges with exactly one endpoint in $C$.
\end{definition}
We prove the following theorem for min-max correlation clustering.
\vspace{-0.1cm}
\begin{restatable}{thm}{minMaxCC}
\label{min-max-CC}
Given an edge weighted graph $G=(V,E)$ on $n$ vertices such that each edge is labeled positive or negative, there exists a polynomial time algorithm which outputs a clustering $\mathcal{C} = \{C_1,\cdots, C_{|\mathcal{C}|}\}$ of $G$ such that the disagreement on each $C_i\in \mathcal{C}$ is at most $\mathcal{O}(\sqrt{\log n\cdot\max\{\log(|E^-|),\log(k)\}})\cdot OPT$; where $OPT$ is the maximum disagreement on each cluster in an optimal solution of min-max correlation clustering, $k$ is the number of clusters in the optimum solution, and $|E^-|$ denotes the number of negative edges in $G$.
\end{restatable}
\vspace{-0.1cm}
In order to prove Theorem \ref{min-max-CC}, we give a reduction from the problem of min-max correlation clustering to a problem which we call min-max multicut. 
\begin{definition}
(Min-max Multicut)
Given an edge weighted graph $G=(V,E)$ and a set of source-sink pairs $\{(s_1,t_1),\cdots, (s_T,t_T\}$, the goal is to give a partitioning $\mathcal{P} = \{P_1,P_2,\cdots,P_{|\mathcal{P}|}\}$ of $G$ such that all the source sink pairs are separated, and $\max_{1\leq i \leq |\mathcal{P}|}\delta(P_i)$ is minimized.
\end{definition}
In min-max multicut, we do not force each part of the partitioning to have a terminal and there could be some parts without any terminals in the final solution. However, in the min-max multiway cut problem introduced by Svitkina and Tardos~\cite{svitkina2004min}, each part needs to have exactly one terminal.
We prove the following theorem for min-max multicut:
\vspace{-0.2cm}
\begin{restatable}{thm}{minMaxMC}
\label{min-max-MC}
Given an edge weighted graph $G=(V,E)$ on $n$ vertices, and a set of source sink pairs $S_G = \{(s_1,t_1),\cdots,(s_T,t_T)\}$, there exists a polynomial time algorithm which outputs a partitioning $\mathcal{P} = \{P_1,\cdots, P_{|\mathcal{P}|}\}$ of $G$, such that all the source sink pairs are
separated and $\max_{1\leq i \leq |\mathcal{P}|}\delta(P_i)\leq \mathcal{O}(\sqrt{\log n\cdot\max\{\log(T),\log(k)\}})\cdot OPT$; where $OPT$ is the value of the optimum solution of min-max multicut, and $k$ is the number of clusters in the optimum solution.
\end{restatable}
We also consider the following variation of min-max multicut called min-max constrained multicut. In this variation, the goal is to partition a graph into a minimum number of parts to separate all source-sink pairs.
\begin{definition}
(Min-max Constrained Multicut)
An edge weighted graph $G=(V,E)$ and a set of source-sink pairs $\{(s_1,t_1),\cdots, (s_T,t_T)\}$ is given. Given $k$ the minimum number of parts needed to separate all source sink pairs, the goal is to partition $G$ into $k$ parts $\{P_1,\cdots,P_k\}$ which separate all source-sink pairs, and $\max_{1\leq i \leq k}\delta(P_i)$ is minimized. 
\end{definition}
We get improved approximation ratios for min-max correlation clustering, min-max multicut on graphs excluding a fixed minor.
\vspace{-0.2cm}
\begin{restatable}{thm}{MinorCC}
\label{MinorCC}
Given an edge weighted graph $G$ excluding $K_{r,r}$ minors, there exist polynomial time $\mathcal{O}(r^2)$-approximation algorithms for min-max correlation clustering and min-max multicut.
\end{restatable}
\vspace{-0.2cm}
Finally, we get improved approximation ratio for min-max correlation clustering on complete graphs.
\begin{restatable}{thm}{CompleteCC}
\label{thm:min-max-CC}
Given an unweighted complete graph on the set of  vertices $V$ ($|V|=n$) such that each edge is labeled positive or negative, there exists a polynomial time algorithm which outputs a clustering $\mathcal{C} = \{C_1,\cdots, C_{|\mathcal{C}|}\}$ of $G$ such that the disagreement on each $C_i\in \mathcal{C}$ is at most $14\cdot OPT$; where $OPT$ is the maximum disagreement on each cluster in an optimal solution of min-max correlation clustering.
\end{restatable}
\vspace{-0.2cm}

\vspace{-0.2cm}
\subsection{High Level Ideas}
Most algorithms for correlation clustering with the global minimizing disagreement objective use a linear programming relaxation \cite{chawla2015near,demaine2006correlation,charikar2003clustering}. The recent work of Charikar, Gupta and Scharwtz also uses a similar linear programming relaxation for the vertex-wise min-max objective \cite{charikar2017local}. Surprisingly, these relaxations do not work for the min-max correlation clustering problem considered in this paper. Indeed, simply obtaining a linear programming relaxation for the cluster-wise min-max objective looks challenging!

Bansal et al.~\cite{bansal2014min} considered a semidefinite programming (SDP) based approximation algorithm for min-max $k$ balanced partitioning and min-max multiway cut with $k$ terminals. In their approach, instead of finding the entire solution in one shot, they obtain a single part at a time. It is possible to encode the same problem with a linear program albeit with a worse approximation guarantee. They use SDP rounding to obtain a part with low cut capacity, and repeat the process until the parts produce a covering of all the vertices. By properly adjusting the weight of each part, the covering can be obtained efficiently. Finally, they convert the covering to partitioning.

The problem of extracting a single cluster of min-max correlation clustering can be captured by a semidefinite programming formulation. Here it is not over a cut capacity objective, instead we need to simultaneously consider the  intra-cluster negative edges as well as inter-cluster positive edges. Indeed, even for the global minimization objective, we are not aware of any good rounding algorithm based on SDP relaxation of correlation clustering. Therefore, rounding the SDP formulation directly looks difficult. To overcome this, we instead consider a new problem of {\em min-max multicut}. Demaine et al.~\cite{demaine2006correlation} have shown an approximation preserving reduction between multicut and correlation clustering (for the global objective function). By solving the min-max multicut problem and then using the aforementioned reduction, we solve the min-max correlation clustering problem. 

First, the reduction of Demaine et al.~\cite{demaine2006correlation} is for the global objective, and an equivalence in global objective does not necessarily correspond to equivalency in local min-max objective. Fortunately, we could show indeed such an equivalency can be proven (Section \ref{cluster-wise-CC}). Thus, the ``multicut'' route seems promising as it optimizes over a cut objective. We consider obtaining each component of the min-max multicut problem, repeat this process to obtain a covering \cite{bansal2014min}, and finally convert the covering to a partitioning. 

The major technical challenge comes in rounding the SDP relaxation for the multicut instance where we seek to find a single component with good cut property. In order for the relaxation to be valid, we have to add new constraints so that no source-sink pair $(s_i,t_i)$ appears together. We also need to ensure that the component obtained satisfies a weight lower bound by assigning weights to each vertex. This is important in the next step when we wish to get a covering of all the vertices: we will decrease the weight of the vertices in the component recovered and again recompute the SDP relaxation with the same weight lower bound. This ensures the same component is not repeatedly recovered and a final covering can be obtained. 
To solve min-max multiway cut, Bansal et al.~\cite{bansal2014min} need to separate $k$ terminals. To do so, they can just guess which of the $k$ terminals if any should appear in the current component with only $k+1$ guesses. For us, the number of such guesses would be $3^{T}$ where $T$ is the number of source sink pairs since for every pair $(s_i,t_i)$, either $s_i$ or $t_i$ or none would be part of the returned component. Since $T$ could be $O(n^2)$ such a guessing is prohibitive. We need to come up with a new approach to address this issue. 

We use a SDP relaxation to compute a metric on the graph vertices and add additional constraints to separate source sink pairs along with the spreading constraints from Bansal et al.~\cite{bansal2014min} to recover a component of desired size. Next, we use the SDP separator technique introduced by Bansal et al.~\cite{bansal2014min} to design a rounding algorithm that returns a set $S = \{S_1, S_2, \cdots, S_j\}$, such that for each $S_i\in S$, there are no source-sink pairs in $S_i$. Bansal et al.~\cite{bansal2014min} need to glue the sets in $S$ and report it as a single component, since they wish to get a solution with specified number of components at the end. However, in min-max multicut problem, the number of components does not matter. Therefore, we do not need to union the sets in $S$, and as a result no source-sink violations happen.


It is possible to use a linear programming formulation for the detour via multicut and use LP-separators of Bansal et al.~\cite{bansal2014min} in place of orthogonal separators and follow our algorithm. This would achieve a similar bound for min-max multicut and min-max correlation clustering in general graphs, but a much better bound of $\mathcal{O}(r^2\cdot OPT)$ for graphs that exclude $K_{r,r}$ minors. 
Similarly, we use LP formulation of correlation clustering problem to devise a new algorithm for complete graphs. 
\vspace{-0.5cm}

\section{Min-Max Multicut}
\vspace{-0.3cm}
Given a subset $S\subseteq V$, let $\delta(S)$ denote the number of edges with exactly one end-point in $S$ and let the number of source sink pairs $(s_i,t_i)$ such that both $s_i$ and $t_i$ belong to $S$ be $vio(S)$.

In order to prove Theorem \ref{min-max-MC}, we first wish to find a set $S=\{S_1,\cdots,S_j\}$, such that $\forall S_i\in S, S_i\subseteq V$, and $\delta(S_i)\leq \mathcal{O}(\sqrt{\log n\cdot\max\{\log(|T|),\log(k)\}})\cdot OPT$, where $OPT$ is the maximum number of cut edges on each part of the optimum partitioning for the min-max multicut problem on graph $G$, $k$ is the number of clusters in the optimum solution which is guessed, $T$ is the number of source-sink pairs, and $n$ is the number of vertices in $G$. In addition, $\forall S_i\in S,vio(S_i)=\mathcal{O}(1)$.

Graph $G=(V,E)$ can have arbitrary edge weights, $w: E \rightarrow \mathbb{R}^+$.
We assume graph $G=(V,E)$ is also a vertex-weighted graph, and there is a measure $\eta$ on $V$ such that $\eta(V)=1$. This measure is used to get a covering of all the vertices. In Section \ref{covering-aggregation-section}, Theorem \ref{thm-find-one-set-multicut} is repeatedly applied to generate a family of sets that cover all the vertices. When a vertex is covered its weight is decreased so the uncovered vertices have a higher weight. Constraint $\eta(S)\in $ $\eta(S)=\sum_{i=1}^j\eta(S_i) \in \big[H/4, 12H\big]$ makes sure the newly computed family of sets $S$ has adequate coverage.
Parameter $H\in(0,1)$ is equal to $1/k$ where $k$ is the number of parts in the optimum partitioning which we guess. 

After getting a covering of all the vertices, in Section \ref{covering-aggregation-section}, it is explained how to convert a covering into a partitioning with the properties desired in Theorem \ref{min-max-MC}.
In order to prove Theorem \ref{min-max-CC}, 
in the full version of this paper
we show how a $\mathcal{O}(\sqrt{\log n\cdot\max\{\log(|T|),\log(k)\}})$-approximation algorithm for min-max multicut implies a $\mathcal{O}(\sqrt{\log n\cdot\max\{\log(|E^-|),\log(k)\}})$-approximation algorithm for min-max correlation clustering.

First we prove the following theorem:
\begin{thm}
\label{thm-find-one-set-multicut}
We are given an edge-weighted graph $G = (V,w)$, a set of source sink pairs $S_G$, a measure $\eta$ on $V$ such that $\eta(V) = 1$, and a parameter $H\in(0,1)$. Assume there exists a set $S^*\subseteq V$ such that $\eta(S^*)\in [H,2H]$, and $vio(S^*)=0$. We design an efficient randomized algorithm to find a set $S$, where $S=\{S_1,\cdots, S_j\}$ satisfying $\forall S_i\in S, S_i \subseteq V$, $\eta(S)= \sum_{i=1}^j \eta(S_i) \in \big[H/4, 12H\big]$, and $\forall S_i \in S$, $vio(S_i) = 0$, and:
\vspace{-0.2cm}
\[\delta(S_i) \leq \mathcal{O}(\sqrt{\log n\cdot\max\{\log(T),\log(k)\}})\cdot OPT\]
where $OPT = \argmin \big\{\delta(S^*) : \eta(S^*)\in [H,2H], \forall (s_i,t_i)\in S_G, |\{s_i,t_i\}\cap S^*|\leq 1\}$ and $|S_G| = T$.
\end{thm}
In order to prove this theorem, we use the notion of $m-$orthogonal separators, a distribution over subsets of vectors, introduced by Chlamtac et al. \cite{chlamtac2006play} which is explained in the following:
\begin{definition}
Let $X$ be an $\ell_2^2$ space (i.e a finite collection of vectors satisfying $\ell_2^2$ triangle inequalities with the zero vector) and $m>0$. A distribution over subsets $S$ of $X$ is an $m-$orthogonal separator of $X$ with probability scale $\alpha>0$, separation threshold $0<\beta<1$, and distortion $D>0$, if the following conditions hold:
\begin{itemize}
\item $\forall u\in X, \Pr(u \in S) = \alpha \norm{u}^2$
\item $\forall u,v \in X \text{ if } \norm{u-v}^2\geq \beta\min\{\norm{u}^2 , \norm{v}^2\} \text{ then }\Pr(u\in S \text{ and } v\in S)\leq \frac{\min\{\Pr(u\in S), \Pr(v\in S)\}}{m}$
\item $\forall u,v \in X$, $\Pr(I_S(u)\neq I_S(v)) \leq \alpha D \cdot \norm{u-v}^2$, where $I_S$ is the indicator function for the set $S$.
\end{itemize}
\end{definition}
Operator $\norm{.}$ shows the $\ell^2$ norm.
Chlamtac et al. \cite{chlamtac2006play} proposed an algorithm for finding $m$-orthogonal separators.
\vspace{-0.2cm}
\begin{thm}
\cite{chlamtac2006play}
There exists a polynomial-time randomized algorithm that given an $\ell_2^2$ space $X$ containing $0$ and a parameter $m>0$, and $0<\beta<1$, generates an $m-$orthogonal separator with distortion $D = \mathcal{O}_\beta(\sqrt{\log|X|\log m})$ and $\alpha \geq \frac{1}{poly(|X|)}$.
\end{thm}
\vspace{-0.5cm}
\subsection{SDP Relaxation}\label{sec:sdp}
In order to prove Theorem \ref{thm-find-one-set-multicut}, we use the following SDP relaxation which is inspired by Bansal et al.~\cite{bansal2014min} except for Constraints \ref{src-sink-constraint-1} and \ref{src-sink-constraint-2}. 
In this relaxation, we assign a vector $\bar{v}$ for each vertex $v\in V$. The objective is to minimize the total weight of cut edges. The set of Constraints \ref{triangle-inequality} are $\ell_2^2$ triangle inqualities, and the set of Constraints \ref{triangle-inequality-zero-vector1} and \ref{triangle-inequality-zero-vector2} are $\ell_2^2$ triangle inequalities with the zero vector. The set of Constraints \ref{src-sink-constraint-1} and \ref{src-sink-constraint-2} make sure that for each source-sink pair $(s_i,t_i)$, both $s_i$ and $t_i$ do not belong to $S$ since both vectors $\bar{s_i}$ and $\bar{t_i}$ could not be \textbf{1} for some fixed unit vector simultaneously.
Constraint \ref{H-constraint-1} and the set of Constraints \ref{H-constraint-2} make sure the returned subgraph has the desired size. 
Suppose now that we have approximately guessed the measure $H$ of the optimal solution $H\leq \eta(S)\leq 2H$. 
We can ignore all vertices $v\in V$ with $\eta(v)>2H$ since they do not participate in the optimal solution and thus write the set of Constraints \ref{H-constraint-2}. Constraints (\ref{spreading-constraint}) are spreading constraints introduced by Bansal et al.~\cite{bansal2014min} which ensure size of $S$ is small.
\vspace{-0.3cm}
\begin{figure}[H]
\setlength{\fboxsep}{1pt}
\setlength{\abovecaptionskip}{1pt}
\begin{center}
\vspace{-0.4cm}
\fbox{
\begin{minipage}[c]{\textwidth}
\vspace{-0.3cm}
\begin{align}
&\min \textstyle\sum\nolimits_{(u,v)\in E} w(u,v)\norm{\bar{u}-\bar{v}}^2\\
\label{triangle-inequality}&\norm{\bar{u}-\bar{w}}^2 + \norm{\bar{w}-\bar{v}}^2 \geq \norm{\bar{u}-\bar{v}}^2 && \forall u,v,w \in V\\
\label{triangle-inequality-zero-vector1}&\norm{\bar{u}-\bar{w}}^2 \geq \norm{\bar{u}}^2 - \norm{\bar{w}}^2 && \forall u,w \in V\\
\label{triangle-inequality-zero-vector2}&\norm{\bar{u}}^2 + \norm{\bar{v}}^2 \geq \norm{\bar{u}-\bar{v}}^2 && \forall u,v \in V\\
\label{src-sink-constraint-1}&\norm{\bar{s_i} - \bar{t_i}}^2 \geq \norm{\bar{s_i}}^2 && \forall (s_i,t_i)\in S_G\\
\label{src-sink-constraint-2}&\norm{\bar{s_i} - \bar{t_i}}^2 \geq \norm{\bar{t_i}}^2 && \forall (s_i,t_i)\in S_G\\
\label{H-constraint-1}&\textstyle\sum\nolimits_{v\in V} \norm{\bar{v}}^2\eta(v) \geq H\\
\label{H-constraint-2}&\norm{\bar{v}}^2 = 0 && \text{if } \eta(v)>2H\\
\label{spreading-constraint}&\textstyle\sum\nolimits_{v \in V}\eta(v)\cdot \min\{\norm{\bar{u}-\bar{v}}^2, \norm{\bar{u}}^2\} \geq (1-2H)\norm{\bar{u}}^2 && \forall u\in V
\end{align}
\end{minipage}
}
\label{fig:SDP}
\end{center}
 \vspace{-0.9cm}
\end{figure}
\vspace{-0.5cm}
\begin{lem}
\label{SDP-relaxation-lemma}
Given $S^* = \argmin \big\{\delta(T) : \eta(T)\in [H,2H], \forall (s_i,t_i)\in S_G, |\{s_i,t_i\}\cap T|\leq 1\}$, the optimal value of SDP is at most 
$\delta(S^*)$.
\end{lem}
\label{proof-SDP-relaxation-appendix}
\begin{proof}
Consider the following solution for the SDP. For each vertex $v$, if $v\in S^*$ let $\bar{v} = \textbf{1}$ for some fixed unit vector, and $\bar{v} = \textbf{0}$ otherwise. We show this gives a feasible solution for the SDP. Clearly triangle inequalities hold for this solution. Now we show that Constraints \ref{src-sink-constraint-1} hold for $S^*$. Consider a source-sink pair $(s_i,t_i)$. Two cases might happen, either $\bar{s_i} = \textbf{0}$ or $\bar{t_i} = \textbf{0}$. If $\bar{s_i} = \textbf{1}$ then $\bar{t_i}$ should be zero since $s_i$ and $t_i$ could not both belong to $S^*$ and the constraint holds. If $\bar{s_i}=\textbf{0}$ then RHS is $0$ and the constraint holds. A similar argument shows the set of Constraints \ref{src-sink-constraint-2} hold as well.
$H$ was guessed such that $H\leq \eta(S^*)\leq 2H$ therefore Constraints \ref{H-constraint-1} and Constraints \ref{H-constraint-2} hold.
Now we want to show the set of Constraints (\ref{spreading-constraint}) hold. Consider a fixed vertex $u$. If $\bar{u}=\textbf{0}$ then both sides of the spreading constraint become $0$. If $\bar{u}=\textbf{1}$ the spreading constraint equals $\eta(V\setminus S^*)\geq (1-2H)$ which holds since $H\leq \eta(S^*)\leq 2H$.
Therefore $S^*$ is a feasible solution and the objective value of SDP is at most $\delta(S^*)$.
\end{proof}
\vspace{-0.5cm}
\subsection {Approximation Algorithm}
\vspace{-0.2cm}
\label{approx-alg}
In this section, we prove Theorem \ref{thm-find-one-set-multicut}. We propose an approximation algorithm which is inspired by Bansal et al.'s ~\cite{bansal2014min} algorithm for small-set expansion (SSE). However, there is a significant difference between our algorithm and theirs. In the SSE problem, one does not need to worry about separating source sink pairs. 

First, we solve the SDP relaxation, and then proceed iteratively. In each iteration, we sample an ($32T\cdot k$) - orthogonal separator $S$ with $\beta = 1/2$ and return it (we repeatedly sample $S$, until a particular function\footnote{defined later} $f(S)$ has some
positive value. Details are deferred to Section \ref{analysis}). Then, $S$ is removed from graph $G$ and the SDP solution, by zeroing the weight of edges incident on $S$ (i.e discarding these edges), and zeroing the SDP variables corresponding to vertices in $S$.
The algorithm maintains the subsets of vertices removed so far in a set $U\subseteq V$, by initializing $U = \emptyset$, and then at each iteration by updating $U = U\cup \{S\}$. We keep iterating until $\eta(U) = \sum_{S_i \in U} \eta(S_i) \geq H/4$. After the last iteration, if $\eta(U)>H$, we output $F=S$ where $S$ is computed in the last iteration. Otherwise, we put $F=U$. Note that in this case, $U = \{S_1,\cdots,S_{|U|}\}$.
\vspace{-0.2cm}
\subsection{Analysis}
\label{analysis}
First, let's see what is the effect of algorithm's changes to the SDP solution. By zeroing vectors in $S$ and discarding the edges incident on $S$, the SDP value may only decrease.
Triangle inequalities, and the source-sink constraints still hold.
Constraint $\sum_{v\in V} \norm{\bar{v}}^2\eta(v) \geq H$ will be violated due to zeroing some variables. However, since before the last iteration $\eta(U)\leq \frac{H}{4}$, the following constraint still holds:
\begin{align}
\label{weaker-weight-constraint}
\textstyle\sum\nolimits_{v\in V} \norm{\bar{v}}^2\eta(v) \geq \frac{3H}{4}
\end{align}
Next, we show the set of spreading constraints (\ref{spreading-constraint}) will remain satisfied after removing $S$. Consider the spreading constraint for a fixed vertex $u$, two cases might happen:

\textbf{Case 1:} If $\exists S\in U$ such that $u\in S$, then $u$ will be removed and $\norm{\bar{u}} = 0$, the spreading constraint will be satisfied since RHS is $0$.

\textbf{Case 2:} If $\nexists S\in U$ such that $u\in S$, the RHS will not change and we can show that $\min\{\norm{\bar{u}-\bar{v}}^2, \norm{\bar{u}}^2\}$ does not decrease. If $\nexists S'\in U$ such that $v\in S'$, then the term $\min\{\norm{\bar{u}-\bar{v}}^2, \norm{\bar{u}}^2\}$ does not change. If $\exists S'\in U$ such that $v\in S'$, then $\min\{\norm{\bar{u}-\bar{v}}^2, \norm{\bar{u}}^2\} = \norm{\bar{u}}^2$ since $\norm{\bar{v}} = 0$, and its value does not decrease.

Therefore, in both these cases, the spreading constraints will not be violated.
\vspace{-0.5cm}
\begin{lem}
\label{lem-bound-eta-S}
Let $S$ be a sampled $(T\cdot k)$-orthogonal separator. Fix a vertex $u$. We
claim that $\Pr[\eta(S)\leq 12H\given u\in S]\geq \frac{7}{8}$.
\end{lem}
\label{appendix-proof-lem-bound-eta-S}
\begin{proof}
Consider a vertex $u$ and let $A_u = \{v: \norm{\bar{u}-\bar{v}}^2 \geq \beta \norm{\bar{u}}^2\}$ and $B_u = \{v: \norm{\bar{u}-\bar{v}}^2 < \beta \norm{\bar{u}}^2\}$. Assume for now that $u\in S$. We show with high probability $\eta(A_u \cap S)$ is small, and $\eta(B_u)$ is also small. 
Vertex $u$ satisfies the spreading constraint. It is
easy to see that:
\[(1-2H)\norm{u}^2 \leq \textstyle\sum_{v \in V}\eta(v)\cdot \min\{\norm{\bar{u}-\bar{v}}^2, \norm{\bar{u}}^2\} \leq \beta \norm{\bar{u}}^2 \eta(B_u) + \norm{\bar{u}}^2 \eta(A_u)\]
Since $\eta(V) = 1$ and $A_u \cup B_u = V$,
$\eta(A_u)+\eta(B_u) = 1$, and $\beta=1/2$ therefore:\\
\begin{align}
& (1-2H) \leq \beta \eta(B_u) + (1-\eta(B_u))\\
\label{eta-B-upper-bound}&\therefore \eta(B_u) \leq \frac{2H}{1-\beta} = 4H
\end{align}
Consider an arbitrary vertex $v\in A_u$ where $\norm{\bar{v}}\neq 0$. By definition of $A_u$, $\norm{\bar{u}-\bar{v}}^2 \geq \beta \norm{\bar{u}}^2 \geq \beta \min \{\norm{\bar{u}}^2, \norm{\bar{v}}^2\}$.
Therefore, by the second property of orthogonal separators and since we assumed $u\in S$, then $\Pr[v\in S \given u \in S] \leq \frac{1}{32Tk}\leq \frac{1}{32k} \leq H$.

Now we show a bound for $\E[\eta(A_u\cap S)\given u\in S]$:
\[\E[\eta(A_u\cap S)\given u\in S] = \textstyle\sum_{v\in A_u}\eta(v)\Pr[v\in S \given u \in S] \leq H\]
Now, we want to bound $\Pr[\eta(S)\geq 12H\given u\in S]$. The event $\{\eta(S)\geq 12H\given u\in S\}$ implies the event $\{\eta(A_u\cap S)\geq 8H\given u\in S\}$ since $\eta(B_u\cap S)\leq \eta(B_u)\leq 4H$. (The second inequality holds by (\ref{eta-B-upper-bound})).
Now we are ready to complete the proof.
\[\Pr[\eta(S)\geq 12H\given u\in S] \leq \Pr [\eta(A_u\cap S)\geq 8H\given u\in S]
\leq \frac{\E[\eta(A_u\cap S)\given u\in S]}{8H} \leq \frac{H}{8H} = 1/8\]
We showed $\Pr[\eta(S)\geq 12H\given u\in S]\leq 1/8$, therefore $\Pr[\eta(S)\leq 12H\given u\in S]\geq 7/8$ and the proof is complete.
\end{proof}
Next, we upper bound $\delta(S)$. By the third property of orthogonal separators:
\[\E[\delta(S)]\leq \alpha D\cdot\textstyle\sum\nolimits_{(u,v)\in E}\norm{\bar{u}-\bar{v}}^2 \cdot w(u,v)\leq \alpha D\cdot SDP\]
Where $D =\mathcal{O}_{\beta}(\sqrt{\log n\cdot\log(32T\cdot k)}) = \mathcal{O}(\sqrt{\log n\cdot\max\{\log(T),\log(k)\}})$. Note that $\beta = 1/2$.
Consider the function $f$:
\[
  f(S) =
  \begin{cases}
  \eta(S)-\delta(S)\cdot \frac{H}{32D\cdot SDP} - 16vio(S)\cdot Hk & \text{if $S\neq \emptyset$ and $\eta(S)< 12H$}\\
  0 & \text{otherwise}
  \end{cases}
\]
We wish to lower bound $\E[f(S)]$. First, we lower bound $\E[\eta(S)]$. As a result of Lemma \ref{lem-bound-eta-S} and the first property of orthogonal separators:
\begin{align*}
&\E[\eta(S)] = \textstyle\sum\nolimits_{u\in V}\Pr[u\in S \wedge \eta(S)<12H]\cdot \eta(u)\\
&= \textstyle\sum_{u\in V}\Pr[\eta(S) < 12H \given u\in S]\cdot \Pr[u\in S]\cdot \eta(u)
\geq \textstyle\sum\nolimits_{u\in V}\frac{7\alpha\norm{\bar{u}}^2 \eta(u)}{8}
\end{align*}
In the following we put a bound on $\E[vio(S)]$:
\[\E[vio(S)] = \sum_{1\leq i \leq T}\textbf{1}(s_i\in S \wedge t_i \in S)\leq \sum_{1\leq i \leq T}\frac{\alpha\min\{\norm{\bar{s_i}}^2, \norm{\bar{t_i}}^2\}}{32Tk}\leq \frac{\alpha T}{32Tk} = \frac{\alpha}{32k}\]
Since $\E[vio(S)] \leq \frac{\alpha}{32k}$, $\E[\delta(S)]\leq \alpha D\cdot SDP$ and using Constraint \ref{weaker-weight-constraint}:
\[
\E[f(S)]\geq \textstyle\sum\nolimits_{u\in V}\frac{7\alpha\norm{\bar{u}}^2 \eta(u)}{8} - \alpha \cdot D\cdot SDP \cdot\frac{H}{32D\cdot SDP}-\frac{\alpha}{32k}\cdot 16 Hk \geq \frac{7\alpha\frac{3H}{4}}{8} - \frac{\alpha H}{32}-\frac{\alpha H}{2} = \frac{1}{8}\alpha H 
\]
We have $f(S)\leq 2nH$ since $\norm{\bar{u}} = 0$ whenever $\eta(u)>2H$. Therefore, $\Pr[f(S)>0]\geq \frac{\frac{1}{8}\alpha H}{2nH} = \Omega(\frac{\alpha}{n})$. So after $\mathcal{O}(n^2/\alpha)$ samples, with probability exponentially close to 1, the algorithm finds a set S with $f(S)>0$. If $f(S)>0$ then $\eta(S)\geq \delta(S)\cdot \frac{H}{32D\cdot SDP}$, therefore $\delta(S)\leq \frac{32D\cdot SDP\cdot \eta(S)}{H}$.

Additionally, $f(S)>0$ implies $vio(S)\leq \frac{\eta(S)}{16Hk}< \frac{12}{16k} = \frac{3}{4k}$. The second inequality holds since $\eta(S)<12H$. Since $k\ge 1$, $vio(S)<1$ and hence none of the $(s_i,t_i)$ pairs belong to the same cluster $S$.

Consider the two possible cases for the output $F$:

\textbf{Case 1:} $F = U = \{S_1,S_2,\cdots, S_{|U|}\}$, and $\eta(F) = \sum_{i=1}^{|U|} \eta(S_i)$.
In this case, $\frac{H}{4}\leq \eta(F)\leq H$.
The set $U$ is a set of orthogonal separators and each $S_i \in U$ forms a separate part.

\textbf{Case 2:} $F = S$. In this case, let's show the last iteration of step 1 as $U = U_{old}\cup \{S\}$. We know $\eta(U)>H$, and $\eta(U_{old})<\frac{H}{4}$, therefore $\eta(S)> 3H/4$. Also $f(S)>0$ implies $\eta(S)\leq 12H$. Therefore, $3H/4 < \eta(S)\leq 12H$.

In both cases, $\frac{H}{4}\leq \eta(F)\leq 12H$.

We showed when a set $S_i \in U$ is sampled, $\delta(S_i)\leq \frac{32D\cdot SDP\cdot \eta(S_i)}{H}$. However, in the LHS of this inequality, edges like $(u,v)$ where $u\in S_j, v\in S_i$ and $j<i$ are not considered. We can show $ \sum_{j=1}^{i-1}\delta(S_j,S_i)\leq \sum_{j=1}^{i-1}\frac{32D\cdot SDP\cdot \eta(S_j)}{H}\leq 32D\cdot SDP$ since $\sum_{j=1}^{i-1}\eta(S_j)\leq H$.
Therefore, $\delta(S_i)\leq \frac{32D\cdot SDP\cdot \eta(S_i)}{H} + \sum_{j=1}^{i-1}\delta(S_j, S_i)\leq \mathcal{O}(D\cdot SDP)$ since $\eta(S_i)\leq 12H$.


This completes the proof of Theorem \ref{thm-find-one-set-multicut}.

The following corollary is implied from Theorem \ref{thm-find-one-set-multicut} and is used in the next section.
\vspace{-0.1cm}
\begin{corollary}
\label{corr-find-one-set-multicut}
Given an edge-weighted graph $G=(V,w)$, a set of source sink pairs $S_G$, a measure $\eta$ on $V$ such that $\eta(V)=1$, and a parameter $\tau$, a set $S=\{S_1,\cdots,S_j\}$ could be found satisfying $\forall S_i\in S, S_i\subseteq V, vio(S_i)=0$, and $\delta(S_i)\leq \mathcal{O}(\sqrt{\log n\cdot\max\{\log(T),\log(k)\}})\cdot OPT$, where  $OPT = \argmin\{\delta(S^*): \frac{\eta(S^*)}{\eta(V)}\geq \tau, vio(S^*)=0\}$. In addition, $\eta(S)=\sum_{i=1}^j\eta(S_i)\geq \Omega(\tau\cdot \eta(V))$.
\end{corollary}
\begin{proof}
The algorithm guesses $H\geq \tau$ such that $H\leq \eta(OPT)\leq 2H$.  Guessing is feasible since $0\leq \eta(OPT)\leq n\cdot \eta(u)$, where $u$ is the weight of the heaviest element in $OPT$, and $H$ can be chosen from the set $\{2^t\eta(u): u\in V, t=0,\cdots, \log(n)\}$ of size $\mathcal{O}(n\log(n))$. Theorem \ref{thm-find-one-set-multicut} is invoked with parameter $H$. The obtained solution $S$ satisfies the properties of this corollary. To be more specific, by invoking Theorem \ref{thm-find-one-set-multicut}, $\eta(S)=\sum_{i=1}^j\eta(S_i)\geq \frac{\tau}{4}\cdot \eta(V)$.
\end{proof}

\vspace{-0.3cm}
\subsection{Covering \& Aggregation}
\label{covering-aggregation-section}
Once we find $F$, we follow the multiplicative update algorithm of \cite{bansal2014min} with some minor modifications, to get a covering of all the vertices. Then, we use the aggregation step to convert the covering to a partitioning. This step is simpler than \cite{bansal2014min} since we are not required to maintain any size bound on the subgraphs returned after aggregation. 
\begin{thm}
\label{thm-covering}
Given graph $G=(V,E)$ and $T$ pairs of source and sink, running Algorithm \ref{Covering-Alg} on this instance outputs a multiset $\mathcal{S}$ that satisfies the following conditions:
\begin{itemize}
\item for all $S\in \mathcal{S}$: $\delta(S)\leq D\cdot OPT \text{ where } D= \mathcal{O}(\sqrt{\log n\cdot\max\{\log(T),\log(k)\}}), vio(S) = 0$.
\item for all $v\in V$, $\frac{|\{S\in \mathcal{S}: v\in S\}|}{|\mathcal{S}|}\geq \frac{1}{17 kn}$, where $k$ is the number of parts in the optimal solution which we guess.
\end{itemize}
\end{thm}
\begin{algorithm}[H]
\label{Covering-Alg}
\SetAlgoLined
$\text{Set } t=1, S=\emptyset \text{ and } y^1(v)=1 \ \forall v\in V$\;
Guess $k$, which is the number of parts in the optimal solution\;
\While{$\sum_{v\in V} y^t(v)>\frac{1}{n}$}{
	Find set $S^t =\{S_1,\cdots, S_j\}$ 
	using Corollary \ref{corr-find-one-set-multicut}, where $\tau=\frac{1}{k}$ and $\forall v\in V, \eta(v) = y^t(v)/\textstyle\sum_{v\in V}y^t(v)$\;
	$\mathcal{S} = S^t \cup \mathcal{S}$\;
	// Update the weights of the covered vertices\;
	\For{$v\in V$}{
		Set $y^{t+1}(v) = \frac{1}{2}\cdot y^t(v)$ if $\exists S_i \in S^t$ such that $v\in S_i$, and $y^{t+1}(v) = y^t(v)$ otherwise.\;
	}
	Set $t = t+1$\;
}
return $\mathcal{S}$\;
\caption{Covering Procedure for Min-Max Multicut}
\end{algorithm}
\begin{proof}
For an iteration $t$, let $Y^t = \sum_{v\in V}y^t(v)$. Consider the optimal solution $\{S_i^*\}_{i=1}^k$ to the min-max multicut problem. There exists at least a $S_j^* \in \{S_i^*\}_{i=1}^k$ with weight greater than or equal to the average ($y_t(S_j^*)\geq \frac{Y^t}{k}$), $vio(S_j^*)=0$, and $\delta(S_j^*)\leq OPT$. Therefore by Corollary \ref{corr-find-one-set-multicut} where $H=\frac{1}{k}$, a set $S_t = \{S_1,S_2,\cdots, S_j\}$ could be found where $\forall S_i \in S_t$, $\delta(S_i)\leq \mathcal{O}(\sqrt{\log n\cdot\max\{\log(T),\log(k)\}})\cdot OPT$, $vio(S_i)=0$.

Now we show the second property of the theorem holds. Let $\ell$ denote the number of iterations in the while loop. Let $|\{S\in \mathcal{S}: v\in S\}| = N_v$. By the updating rules $y^{\ell+1}(v) = 1/2^{N_v}$. Therefore $\frac{1}{2^{N_v}} = y^{\ell+1}(v) \leq 1/n$, which implies $N_v \geq \log(n)$.
By Corollary \ref{corr-find-one-set-multicut}, $y^t(S^t)\geq \frac{1}{4 k}Y^t$. Therefore:
\vspace{-0.3cm}
\[Y^{t+1} = Y^t - \frac{1}{2}y^t(S^t)\leq (1-\frac{1}{8 k})Y^t\]
Which implies $Y^{\ell} \leq (1-\frac{1}{8 k})^{\ell - 1}Y^1 = (1-\frac{1}{8 k})^{\ell - 1}n$.
Also $Y^{\ell}>1/n$ therefore, $\ell \leq 1+16 k \ln(n) \leq 17 k \log(n)$.
In each iteration $t$, the number of sets in $S_t$ is at most $n$ (since all the sets in $S_t$ are disjoint), therefore $|\mathcal{S}|\leq 17 k n \log(n)$,
and the second property is proved.
\end{proof}

Now the covering of $G$ is converted into a partitioning of $G$ without violating min-max objective by much.
\vspace{-0.1cm}
\begin{thm}
\label{aggregation}
Given a weighted graph $G=(V,E)$, a set of source-sink pairs $(s_1,t_1),\cdots,(s_{T},t_{T})$, and a cover $\mathcal{S}$ consisting of subsets of $V$ such that:
\begin{itemize}
\item $\forall v\in V$, $v$ is covered by at least a fraction $\frac{c}{nk}$ of sets $S\in\mathcal{S}$, where $k$ is the number of partitions of the optimum solution which we guessed in the covering section, and $c=1/17$.
\item $\forall S\in\mathcal{S}$, $\delta(S)\leq B$, $vio(S)=0$.
\end{itemize}
We propose a randomized polynomial time algorithm which outputs a partition $\mathcal{P}$ of $V$ such that $\forall P_i \in \mathcal{P}$, $\delta(P_i)\leq 2B$, and $vio(P_i) = 0$.
\end{thm}
\begin{algorithm}[H]
\label{aggregation-alg}
\SetAlgoLined
Step 1: Sort sets in $\mathcal{S}$ in a random order: $S_1,S_2,\cdots,S_{|\mathcal{S}|}$. Let $P_i = S_i\setminus (\cup_{j<i} S_j)$. \\
Step 2:
\While{There is a set $P_i$ such that $\delta(P_i)>2B$}{
	Set $P_i = S_i$ and for all $j\neq i$, set $P_j = P_j\setminus S_i$\;
}
\caption{Aggregation Procedure For Min-Max Multicut}
\end{algorithm}
\begin{proof}
\label{aggregation-proof-appendix}
A similar proof to the one given by Bansal et al. \cite{bansal2014min} shows after step 2, for each $P_i\in \mathcal{P}$, $\delta(P_i)\leq 2B$.
We start by analyzing Step 1. Observe that after Step 1, the collection of sets $\{P_i\}$ is a partition of $V$ and $P_i\subseteq S_i$ for every $i$. Particularly, $vio(P_i)\leq vio(S_i)$. Note, however, that the bound $\delta(P_i)\leq B$ may be violated for some $i$ since $P_i$ might be a strict subset of $S_i$.\\
We finish the analysis of Step 1 by proving that $\E[\sum_i \delta(P_i)]\leq 2knB/c$. Fix an $i\leq |\mathcal{S}|$ and estimate the expected weight of edges $E(P_i, \cup_{j>i}P_j)$, given that the $i^{th}$ set in the random ordering is $S$. If an edge $(u,v)$ belongs to $E(P_i, \cup_{j>i}P_j)$, then $(u,v)\in E(S_i, V\setminus S_i) = E(S,V\setminus S)$ and both $u,v\notin \cup_{j<i} S_j$. For any edge $(u,v)\in \delta(S)$ (with $u\in S, v\notin S$), $\Pr((u,v)\in E(P_i,\cup_{j>i}P_j)\given S_i = S)\leq \Pr(v\notin \cup_{j<i}S_j\given S_i = S)\leq (1-\frac{c}{nk})^{i-1}$, since $v$ is covered by at least $\frac{c}{nk}$ fraction of sets in $\mathcal{S}$ and is not covered by $S_i = S$. Hence,
\[\E[w(E(P_i, \cup_{j>i}P_j))\given S_i = S] \leq (1-\frac{c}{nk})^{i-1}\delta(S)\leq (1-\frac{c}{nk})^{i-1}B \]
and $\E[w(E(P_i, \cup_{j>i}P_j))\leq (1-\frac{c}{nk})^{i-1}B$. Therefore:
\[\E\big[\textstyle\sum_i \delta(P_i)\big] = 2 \cdot \E\big[\textstyle\sum_i w(E(P_i, \cup_{j>i}P_j))\big] \leq 2 \textstyle\sum_{i=0}^\infty (1-\frac{c}{nk})^i B = 2knB/c\]
Now we want to analyze step 2. Consider potential function $\sum_i \delta(P_i)$, we showed after step 1, $\E\big[\sum_i \delta(P_i)\big] \leq 2knB/c$. 
We prove that this potential function reduces quickly over the iterations of Step 2, thus, Step 2 terminates after a small number of steps. After each iteration of Step 2, the following invariant holds: the collection of sets $\{P_i\}$ is a partition of $V$ and $P_i\subseteq S_i$ for all $i$. Particularly, $vio(P_i)\leq vio(S_i)$. Using an uncrossing argument, we show at every iteration of the while loop in step 2, $\sum_i \delta(P_i)$ decreases by at least $2B$.
\begin{align*}
&\delta(S_i) + \textstyle\sum_{j\neq i} \delta(P_j\setminus S_i)\leq \delta(S_i) + \textstyle\sum_{j\neq i}\Big(\delta(P_j) + w(E(P_j\setminus S_i,S_i)) - w(E(S_i\setminus P_j,P_j))\Big)\\
&\leq \delta(S_i) + \textstyle\sum_{j\neq i}\Big(\delta(P_j)\Big) + w(E(V\setminus S_i,S_i)) - w(E(P_i,V\setminus P_i))\\
&= \textstyle\sum_{j}\Big(\delta(P_j)\Big)+2\delta(S_i)-2\delta(P_i)\leq \textstyle\sum_{j}\Big(\delta(P_j)\Big) - 2B
\end{align*}
The above inequalities use the facts that $P_i\subseteq S_i$ for all $i$ and that all the $P_j$'s are disjoint. The second inequality uses the facts that $\sum_{j\neq i}w(E(P_j\setminus S_i,S_i)) = w(E(V\setminus S_i,S_i))$, and $\sum_{j\neq i} w(E(S_i\setminus P_j,P_j)) \geq w(E(P_i,V\setminus P_i))$, which hold since the collection of sets $\{P_i\}$ is a partition of $V$, and $P_i\subseteq S_i$. In particular, $\sum_{j\neq i} w(E(S_i\setminus P_j,P_j)) \geq w(E(P_i,V\setminus P_i))$ holds since for each edge $e$ if $e\in E(P_i,P_j)$ then $e\in E(S_i\setminus P_j, P_j)$.
The last inequality holds since $\delta(S_i)\leq B $ and $\delta(P_i)>2B$.\\
This proves that the number of iterations of the while loop is polynomially bounded and after step 2, $\delta(P_i)\leq 2B$ for each $P_i$.\\
In addition, since each $P_i$ is a subset of $S_i$, $vio(P_i)\leq vio(S_i)$. Therefore $vio(P_i)=0$.
\end{proof}

\section{Analysis of Algorithm for Min-Max Correlation Clustering}
\label{cluster-wise-CC}
In order to prove Theorem \ref{min-max-CC},  we reduce a correlation clustering instance to a multicut instance. We follow the reduction shown by Demaine et al. \cite{demaine2006correlation}. They proved that the global objective multicut and correlation clustering are equivalent. However, equivalancy of multicut and correlation clustering with respect to global objective does not immediately imply their equivalancy with respect to min-max objective. In the following, first we mention the reduction, and then we show how a $\mathcal{O}(\sqrt{\log n\cdot\max\{\log(|T|),\log(k)\}})$-approximation algorithm for min-max multicut implies a $\mathcal{O}(\sqrt{\log n\cdot\max\{\log(|E^-|),\log(k)\}})$-approximation algorithm for min-max correlation clustering.

Given a graph $G = (V,E)$ which is an instance of correlation clustering, we construct a new
graph $G' = (V',E')$ and a collection of source sink pairs $S_{G'} = \{\big \langle s_i, t_i \big \rangle \}$ as follows:
Initially $V' = V$. For every negative edge $(u,v)\in E^{-}$ with weight $w{(u,v)}$, we add a new vertex $uv$ to $V'$ and a new edge $(u,uv)$ to $E'$ with weight $w(u,v)$. Also we add a source sink pair $(v,uv)$ to $S_{G'}$. For every positive edge $(u,v)\in E^{+}$ with weight $w(u,v)$, we add $(u,v)$ with weight $w(u,v)$ to $E'$. Now we have a multicut instance on $G'$ with source sink pairs $S_{G'}$.
Using Theorem \ref{min-max-MC}, we find a paritioning $\mathcal{P} = \{P_1,P_2,\cdots, P_{|\mathcal{P}|}\}$ of $G'$.
Next, we show how to convert $\mathcal{P}$ into a clustering $\mathcal{C}$ for graph $G$ and prove Theorem \ref{min-max-CC}.

In order to map a partitioning $\mathcal{P}$ into a clustering $\mathcal{C}$ for graph $G$, for each subset $P_i\in \mathcal{P}$, create a cluster $C_i$ and for all $v\in V$, if $v\in P_i$, add $v$ to $C_i$. We show the number of disagreements on each cluster $C_i\in \mathcal{C}$ ($cost(C_i)$) is at most the cut capacity of the corresponding subset $P_i\in \mathcal{P}$ ($\delta(P_i)$). 
Next, we prove this algorithm gives an $\mathcal{O}(\sqrt{\log n\cdot\max\{\log(|E^-|),\log(k)\}})$-approximation algorithm for min-max correlation clustering.

\begin{lemma}
\label{compare-approx-MMMC-MMCC}
For all $P_i\in \mathcal{P}$ and the corresponding cluster $C_i$, $cost(C_i)\leq \delta(P_i)$.
\end{lemma}
\begin{proof}
We show if $C_i$ pays for an edge, then $P_i$ will also pay for that edge. Consider an arbitrary edge $(u,v)$, it could be either positive or negative.\\
\textbf{Case 1:} $(u,v)$ is a positive edge.
In this case if $C_i$ is paying for $(u,v)$, which happens when one of $u$ or $v$ is in $C_i$ and the other one is in $V\setminus C_i$, without loss of generality assume $u\in C_i$ and $v\in V\setminus C_i$, then $u\in P_i$ and $v\in V'\setminus P_i$. Therefore $P_i$ will also pay for $(u,v)$.\\
\textbf{Case 2:} $(u,v)$ is a negative edge. In this case if $C_i$ is paying for $(u,v)$ then $(u,v)$ is trapped inside $C_i$. Consider the corresponding multicut instance. In this instance, $u,v \in P_i$, there is a new vertex $uv$, a new edge $(u,uv)$ and $(v,uv)$ is a source-sink pair which implies $uv\in V'\setminus P_i$. Assume $uv\in P_j$, then the multicut solution pays for edge $(u,uv)$ on both parts $P_i$ and $P_j$. Therefore if $C_i$ pays for a negative edge, the corresponding part in the multicut partitioning will also pay for that edge.
\end{proof}

\begin{lemma}
\label{compare-opt-MMMC-MMCC}
$cost(C^*)\geq \delta(\mathcal{P}^*)$ where $C^*$ is the optimum solution for the min-max correlation clustering on $G$ and $\mathcal{P}^*$ is the optimum solution for the min-max multicut on $G'$.
\end{lemma}
\begin{proof}
We construct a partitioning $\mathcal{P}$ of $G'$ which separates all the source-sink pairs in $G'$, in addition $cost(C^*) = cost(\mathcal{P})$. For each cluster $C^*_i$, construct a set $P_i$ and $\forall v\in C^*_i$, add $v$ to $P_i$. For all $uv\in V'\setminus V$ initially make them singleton clusters. It is easy to see that all source-sink pairs are separated in $\mathcal{P}$.
Also for all positive edges in $G$, $\mathcal{P}$ and $C^*$ are paying the same price. The only difference in the cost of $\mathcal{P}$ and $C^*$ could happen for negative edges. Two cases might happen:
First, consider a negative edge $(u,v)\in C^*_i$. In this case $C^*_i$ is paying for $(u,v)$. In $\mathcal{P}$, $u\in P_i$ and $uv$ is a singleton cluster. Edge $(u,uv)$ is cut and $P_i$ and the singleton cluster $uv$ are paying for it. Therefore $P_i$ and $C^*_i$ are paying the same price $w(u,v)$ for edge $(u,v)$. The singleton cluster $uv$ is also paying the same price $w(u,v)$ for that edge. In addition the singleton cluster $uv$ is not paying for any other edge which means cost of it is at most cost of $C^*_i$.
The other case is when a negative edge $(u,v)$ is between clusters, i.e $u\in C^*_i, v\in C^*_j$. Therefore $C^*$ is not paying for $(u,v)$ but $\mathcal{P}$ is paying for that edge since $(u,uv)$ is cut in $\mathcal{P}$. In this case we move $uv$ into the part $P_i$. By doing that, source-sink pair $(uv,v)$ is still separated since $uv\in P_i, v\in P_j$. Also since $(u,uv)$ is not cut anymore, $\mathcal{P}$ and $C^*$ are paying the same price for edge $(u,v)$.\\
Therefore:
\[cost(C^*) = cost (\mathcal{P})\]
Where $cost(C^*)$ is the maximum number of disagreements on each cluster of $C^*$ and $cost(\mathcal{P})$ is the maximum number of cut edges on each part of $\mathcal{P}$.\\
Also $cost(\mathcal{P})\geq cost(\mathcal{P}^*)$. Therefore $cost(C^*)\geq \delta(\mathcal{P}^*)$ and the proof is complete.
\end{proof}
Now we are ready to prove Theorem \ref{min-max-CC}.
\minMaxCC*
\begin{proof}
Let $C^*$ be the optimum solution for the min-max correlation clustering on $G$, and $\mathcal{P}^*$ be the optimum solution for the min-max multicut on $G'$. 
By Theorem \ref{min-max-MC} we can find a partitioning $\mathcal{P}$ of $G$ such that $cost(\mathcal{P})\leq \mathcal{O}(\sqrt{\log n\cdot\max\{\log(|T|),\log(k)\}})\cdot cost(\mathcal{P^*})$. We convert partitioning $\mathcal{P}$ into a clustering $C$ as it was explained earlier in this section.
Therefore:
\[cost(\mathcal{P}^*) \leq cost(C^*)\leq cost(C)\leq cost(\mathcal{P})\]
The first inequality holds by Lemma \ref{compare-opt-MMMC-MMCC}. The third inequality holds by Lemma \ref{compare-approx-MMMC-MMCC}.
Since $cost(\mathcal{P})\leq \mathcal{O}(\sqrt{\log n\cdot\max\{\log(|T|),\log(k)\}})\cdot cost(\mathcal{P^*})$ and the number of source-sink pairs in the min-max multi-cut is equal to the number of negative edges in the min-max correlation clustering instance,
it could be seen that:\\
\[cost(C)\leq \mathcal{O}(\sqrt{\log n\cdot\max\{\log(|E^-|),\log(k)\}})\cdot cost(C^*)\]

\end{proof}

\section{Min-Max Correlation Clustering, Min-Max Multicut, and Min-Max Constrained Multicut in Minor-Closed Graph Families}
\label{appendix-LP-section}
In this section, we show improved results for min-max correlation clustering, min-max multicut, and min-max constrained multicut in minor-closed graph families. The procedure is almost similar to what we did for general graphs. We wish to solve min-max correlation clustering on a weighted graph $G=(V,E)$ excluding a fixed minor $K_{r,r}$. First, we do the same reduction proposed by Demaine et al.~\cite{demaine2006correlation} that we mentioned in Section \ref{cluster-wise-CC} to get a multicut instance $G'$. In the following, we show $G'$ excludes $K_{r,r}$ minors as well. After that we prove Theorem \ref{one-set-multicut-fixed-minor} which is similar to Theorem \ref{thm-find-one-set-multicut}.
\begin{lem}
\label{lem-exclude-fixed-minor}
If $G$ is excluding a fixed minor $K_{r,r}$ then $G'$ also excludes minor $K_{r,r}$. 
\end{lem}
\begin{proof}
We get $G'$ from $G$ by deleting some edges and adding some new vertices and connecting them to exactly one vertex of $G$. It could be seen if $G$ was excluding minor $K_{r,r}$ after these operations $G'$ will be excluding minor $K_{r,r}$ as well.
\end{proof}
\begin{thm}
\label{one-set-multicut-fixed-minor}
Given an edge-weighted graph $G = (V,w)$ excluding $K_{r,r}$ minors, a set of source sink pairs $S_G$, a measure $\eta$ on $V$ such that $\eta(V) = 1$ and a parameter $H\in(0,1)$, there is an efficient algorithm to find a set $S$, where $S=\{S_1,\cdots, S_j\}$ satisfying $\forall S_i\in S, S_i \subseteq V$, $\eta(S)= \sum_{i=1}^j \eta(S_i) \in \big[H/4, 12H\big]$ and $\forall S_i \in S$, $vio(S_i)=\mathcal{O}(1)$:
\[\delta(S_i) \leq \mathcal{O}(r^2)\cdot \min \big\{\delta(T) : \eta(T)\in [H,2H], \forall (s_i,t_i)\in S_G, |\{s_i,t_i\}\cap T|\leq 1\}\]
\end{thm}
In order to prove Theorem \ref{one-set-multicut-fixed-minor}, we write an LP which is analouge with the SDP we used for general graphs.
We use some ideas Bansal et al.~\cite{bansal2014min} used to write an LP for min-max $k$-partitioning problem in minor-closed graph families. As Bansal et al.~\cite{bansal2014min} explain, for every vertex $u\in V$ we introduce a variable $x(u)$ such that $0\leq x(u)\leq 1$. For every pair of vertices $u,v\in V$ we introduce a variable $z(u,v)=z(v,u)$ taking values in $\big[0,1\big]$. The intended integral solution corresponding to a set $S\subseteq V$ has $x(u)=1$ if $u\in S$ and $x(u)=0$ otherwise; $z(u,v) = |x(u)-x(v)|$. One could think of $x(u)$ as an analogue of $\norm{\bar{u}}^2$ and of $z(u,v)$ as an analouge of $\norm{\bar{u}-\bar{v}}^2$ in the SDP relaxation.
\begin{figure}[htbp]
\setlength{\fboxsep}{1pt}
\setlength{\abovecaptionskip}{1pt}
\begin{center}
\fbox{
\begin{minipage}[c]{\textwidth}
\begin{align*}
&\min \sum_{(u,v)\in E} w(u,v)z(u,v)\\
&z(u,v)+z(v,w)\geq z(u,w) && \forall u,v,w \in V\\
&|x(u)-x(v)|\leq z(u,v) && \forall u,v \in V\\
&x(u)+x(v)\geq z(u,v) && \forall u,v \in V\\
&|x(s_i)-x(t_i)|\geq x(s_i) && \forall \text{ source-sink pair }(s_i,t_i)\\
&|x(s_i)-x(t_i)|\geq x(t_i) && \forall \text{ source-sink pair }(s_i,t_i)\\
&\sum_{v\in V} x(v)\eta(v) \geq H\\
&x(v) = 0 && \text{if } \eta(v)>2H\\
&\sum_{v \in V}\eta(v)\cdot \min\{x(u), z(u,v)\} \geq (1-2H)x(u) && \forall u\in V\\
&x(u),z(u,v)\in [0,1] && \forall u,v \in V
\end{align*}
\end{minipage}
}
\label{fig:LP}
\end{center}
  \vspace{-0.9cm}
\end{figure}
In order to prove Theorem \ref{one-set-multicut-fixed-minor}, we use a notion of LP-separators introduced by Bansal et al.~\cite{bansal2014min}. 
\begin{definition}
(LP separator)
Given a graph $G=(V,E)$ and numbers $\{x(u),z(u,v)\}_{u,v\in V}$, a distribution over subsets $S\subseteq V$ is an LP separator with distortion $D\geq 1$, probability scale $\alpha > 0$ and separation threshold $\beta \in (0,1)$ if:
\begin{itemize}
\item for all $u \in V$, $\Pr(u \in S) = \alpha \cdot x(u)$
\item for all $u,v \in V$ with $z(u,v)\geq \beta \cdot \min\{x(u),x(v)\}$, $\Pr(u\in S \text{ and } v\in S) = 0$
\item for all $(u,v)\in E$ we have $\Pr(I_S(u)\neq I_S(v))\leq \alpha D\cdot z(u,v)$, where $I_S$ is the indicator function for the set $S$.
\end{itemize}
\end{definition}
The following theorem was proved by Bansal et. al \cite{bansal2014min}.
\begin{thm}
\cite{bansal2014min}
Given a graph $G=(V,E)$ that excludes $K_{r,r}$ minors, numbers $\{x(u),z(u,v)\}_{u,v\in V}$ satisfying the first three constraints of LP and parameter $\beta\in (0,1)$, there exists an algorithm which returns an LP seperator with distortion $D= \mathcal{O}(r^2/\beta)$, probability scale $\alpha = \Omega(1/|V|)$ and separation threshold $\beta$.
\end{thm}
By replacing the SDP relaxation with the LP relaxation and the orthogonal separators with LP separators, Theorem \ref{one-set-multicut-fixed-minor} could be proved.
The rest of procedure is same as what we did for general graphs. At the end, Theorem \ref{MinorCC} can be proved.

Bansal et. al~\cite{bansal2014min} showed for genus $g$ graphs, an LP separator with distortion $\mathcal{O}(\log (g))$ can be obtained. By following a similar approach an $\mathcal{O}(\log (g))$-approximation for min-max multicut and min-max correlation clutering on genus $g$ graphs can be obtained.


\section{Min-Max Correlation Clustering on Complete Graphs}

In order to prove Theorem \ref{thm:min-max-CC}, we assume the existence of a measure $\eta$ on $V$ such that $\eta(V)=1$. This measure is used to generate a covering of all the vertices by leveraging Theorem \ref{thm-find-one-cluster-complete} multiple times. When a vertex is covered, the corresponding weight is decreased so that the uncovered vertices get a higher weight (Using the multiplicative algorithm of \cite{bansal2014min}), followed by partitioning. The covering and partitioning algorithms are same as that of general graphs (Section \ref{covering-aggregation-section}). First, we prove the following theorem, followed by the covering and partitioning  algorithm:

\begin{thm}
\label{thm-find-one-cluster-complete}
We are given an unweighted complete graph $G$ on  the set of vertices $V$ ($|V|=n$) such that each edge is labeled positive or negative,  a measure $\eta$ on $V$ such that $\eta(V) = 1$, and a parameter $H\in(0,1)$. Assume there exists a set $T\subseteq V$ such that $\eta(T)\ge H$. We design an efficient algorithm to find a set $\mathcal{S}$, where $\mathcal{S}=\{S_1,\cdots, S_j\}$ satisfying $\forall S_i\in \mathcal{S}, S_i \subseteq V$, $\eta(\cup S_i)\ge H$, and:
\vspace{-0.2cm}
\[cost(S_i) \leq 7 \cdot \min \big\{cost(T) : \eta(T)\ge H\}\]
\end{thm}

To prove Theorem \ref{thm-find-one-cluster-complete}, we use the following integer  linear program (ILP)  that tries to solve for $T$ with minimum $cost(T)$ such that $\eta(T)\ge H$.

\begin{figure}[htbp!]
\setlength{\fboxsep}{1pt}
\setlength{\abovecaptionskip}{1pt}
\begin{center}
\fbox{
\begin{minipage}[c]{\textwidth}
\begin{align}
&\min  \sum_{(u,v)\in E^+} d(u,v) + \sum_{(u,v)\in E^-}{} (\max \{x(u),x(v)\}-d(u,v))\\
&d(u,w) + d(w,v) \ge d(u,v) , \hskip 0.1\textwidth& \forall u,v,w \in V \label{eq:tr1} \\
&|x(u)-x(v)|\le d(u,v),&\forall u,v\in V\label{eq:tr2}\\
& d(u,v)\le x(u)+x(v),&\forall u,v\in V\label{eq:tr3}\\
& x(u) +x(v) + d(u,v)\le 2,\ \forall u,v\in V\label{eq:tr4}\\
& \sum_{v\in V} x(v)\eta(v) \ge H\\
&x(u), d(u,v) \in \{0,1\},& \forall u,v\in V \label{eq:integer}
\end{align}
\end{minipage}
}
\label{fig:LP}
\end{center}
  \vspace{-0.9cm}
\end{figure}

In this LP formulation, every node $u$ has a variable $x(u)$ and every edge has a disagreement $d(u,v)  \ \forall u,v\in V$.  The constraints \ref{eq:tr1} to \ref{eq:tr3} are the triangle inequality constraints and \ref{eq:tr4} ensures that atmost two of the three variables can have the value of 1. The last constraint ensures that $\eta(T)\ge H$.
 
\begin{lem}
\label{LP-relaxation-lemma}
Given $T^* = \argmin \big\{cost(T) : \eta(T)\ge H\}$, the optimal value of Integer LP is at most 
$cost(T^*)$.
\end{lem}
\begin{proof}
Consider a candidate solution, such that $x(u)=1$ if $u\in T^*$ and $0$ otherwise. Hence, $d(u,v)=1$ only when $x(u)=1$ and $x(v)=0$ or vice versa. This variable assignment, satisfies the triangle inequalities and also $\sum_{v\in V}\eta(v) x(v) = \eta(T^*) \ge H$. The contribution of the edges to the objective function is as follows:
\begin{enumerate}
    \item $u,v\in T^*$ implies $x(u)=x(v)=1$ and $d(u,v)=0$. The contribution of (u,v) is 0 if $(u,v)\in E^+$ and $1$ otherwise.
    \item $u,v\notin T^*$ implies $x(u)=x(v)=0$ and $d(u,v)=0$. The contribution of any edge $(u,v)$ when $x(u)=x(v)=0$ is 0.
    \item $u\in T^*,\ v\notin T^*$ implies $x(u)=1$ and $x(v)=0$, hence $d(u,v)=1$. The contribution of $(u,v)$ is 0 if $(u,v)\in E^-$ and $1$ otherwise.
\end{enumerate}
This shows that the objective function captures the number of positive edges within $T^*$ and negative edges to nodes outside $T^*$, which is equal to $cost(T^*)$. Hence, the optimal solution of this integer program has objective value at most  $cost(T^*)$.
\end{proof}

\subsection{Approximation Algorithm}
We consider LP relaxation of the integer program with constraints \ref{eq:integer} modified to $x(u), d(u,v) \in [0,1], \forall u,v\in V$ and use Algorithm \ref{alg:guess} to solve for $T$.
We solve the LP relaxation by guessing a node in the optimal cluster. For every guess $u\in V$, we add a constraint $x(u)=1$ in the above LP relaxation and identify the corresponding optimal fractional solution. Suppose $d_{u}$ and $x_u$ is the corresponding optimal fractional solution with objective value $o_u$ when $u$ is the chosen guess. We sort these objective values in non-decreasing order to get a sorted list $\mathcal{O}=\{o_1,\ldots, o_{|V|}\}$  such that  $o_i$ is the optimal objective value of the LP relaxation when $u_i\in V$ is chosen as a guess. We process the sorted list to identify the smallest index $\lambda$ such that  $\sum_{j<\lambda} \eta(u_j) < H \leq \sum_{j\leq \lambda} \eta(u_j)$  and consider the set of these guesses, $\Gamma = \{u_i, i\leq \lambda\}$. 

Firstly, the objective value $o_j,\ \forall j\le \lambda$ is less than the optimal value of the integral objective function ($o_i\leq OPT, \ i\leq \lambda$; See Lemma \ref{lem:guess}). Secondly, for each guess $u_i\in \Gamma$, we run the rounding Algorithm \ref{alg:greedy} to construct an integer solution ($S_i\supseteq \{u_i\}$) which generates a 7-approximation of the corresponding fractional solution ($cost(S_i) \leq 7 o_i$; See Lemma \ref{lem:seven}). This guarantees that each of the integer solution returned by Algorithm 
\ref{alg:guess} is a 7-approximation of the optimal solution to the integer LP. Additionally, $\Gamma \subseteq \cup S_i$ and $\sum_{i\leq \lambda}\eta(u_i) \ge H$ ensures that the $\eta(\cup S_i) > H$.  This completes the proof of Theorem \ref{thm-find-one-cluster-complete}.

\begin{algorithm}[H]
\SetAlgoLined
\For{$u_i\in V$}{
 Let $o_{u_i},d_{u_i},x_{u_i}$ be the solution on solving the LP relaxation with an additional constraint $x(u_i)=1$}
 Sort $\{o_{u}:u\in V\}$ in non-decreasing order to generate a sorted list: $\{o_1,\ldots, o_{|V|}\}$, where $o_i$ corresponds to the guess $u_i$ \\
 Let $\lambda \leftarrow \min_t : \sum_{i=1}^t \eta(u_i) \ge H $
 and $\Gamma \leftarrow \{u_i: i\leq \lambda\}$\\
 $\mathcal{S}\leftarrow \phi$\\
\For{$u_i\in \Gamma$}{
 $S_i\leftarrow$ Use Algorithm \ref{alg:greedy} to round the LP solution,  $(d_{u_i},x_{u_i})$ \\
 $\mathcal{S}\leftarrow \mathcal{S}\cup \{S_i\}$
}
\Return $\mathcal{S}$
\caption{Generate Covering\label{alg:guess}}
\end{algorithm}

Now, we prove the following lemma's :
\begin{lem}
For every guess $u_j$, $j\leq \lambda$, the optimal solution of the LP relaxation $o_j\leq OPT$, where $OPT$ is the optimal integral solution of the integer program considered.\label{lem:guess}
\end{lem}
\begin{proof}
Let $C$ denote the optimal integral solution of the Integer LP i.e. $x(v)=1,\ \forall v\in C$ and $0$ otherwise. Consider the LP relaxation when $u_i$,  $i\leq \lambda$ is guessed. If $u_i\in C$, then $C$ is a valid solution to the LP relaxation. Hence the objective value of the LP relaxation, $o_i\leq OPT$. 

Suppose $\exists i\leq\lambda$ such that  $u_i\notin C $. In this case, $\sum_{j<i} \eta (u_j) < H$ because $i\leq \lambda$. Hence there must exist $k$ such that $k>i$ and $u_k\in C$, because $\sum_{u\in C} \eta(u) \geq H$. Since, the objective values $o_i$'s are arranged in non-decreasing order of objective value, $o_i\leq o_k$ and since $u_k\in C$, $o_k\leq OPT$. Hence $o_i\leq OPT$.
\end{proof}

For every  guess $u\in \Gamma$, we show that the  cluster returned by the rounding Algorithm \ref{alg:greedy} is 7-approximation of the optimal objective value of  the corresponding  LP relaxation. Hence, we will get a candidate solution for each  guess. Below, we show the rounding algorithm and the corresponding approximation ratio.


\subsection{Rounding Algorithm for a particular guess}
Our rounding algorithm is motivated by the ball growing approach in \cite{charikar2003clustering}. We consider a ball of radius $2/7$ (say $T$) around the guessed vertex and try to construct a cluster based on the total fractional disagreements of the vertices in $T$. If the total fractional disagreements are larger than $1/7$ fraction of the number of vertices in the ball, it outputs a singleton cluster with the guessed vertex. On the other hand, if the total disagreements are lower than $1/7$ fraction, it outputs the complete ball $T$ along with the guess.

\begin{algorithm}[H]
\SetAlgoLined
 $T=\{w\in V-\{u\}: d(u,w)\leq \frac{2}{7}\}$\\
\If{$\sum_{w\in T} d(u,w)\ge  |T|/7$}{
 Output the cluster $\{u\}$}
\Else{
 Output the cluster $\{u\}\cup T$
 }

\caption{Rounding Algorithm for a guess $u$\label{alg:greedy}}
\end{algorithm}

\begin{lem}
Algorithm \ref{alg:greedy} identifies a cluster $C$ such that the integral disagreements of $C$ is 7-approximation of the corresponding fractional disagreements.\label{lem:seven}
\end{lem}
\begin{proof}
We consider  two different cases based on the output of the algorithm. For each case, we show that the integral contribution of an edge (or a combination of edges) is less than $7$ times the fractional contribution of the corresponding edge (or corresponding combination of edges).

 Notice that, constraint \ref{eq:tr2} implies that $x(u)-x(v) = 1-x(v)\le d(u,v)$ and, constraint \ref{eq:tr4} implies $x(v)\le 2- x(u)-d(u,v)=1-d(u,v)$ hence $x(v)= 1-d(u,v)\geq 1-2/7 = 5/7 $\\
\noindent \textbf{Case 1:} Only the node $u$ is output as the cluster.
In this case, the integral contribution to the objective is the set of positive neighbors of $u$. Consider the edge $(u,v)$ such that $d(u,v)>2/7$. In this case, the integral contribution is less than $7/2$ times the fractional disagreement of that edge. When $d(u,v)\le 2/7$, the integral contribution of those edges is atmost $|T|$. Also, since  $\sum_{w\in T} d(u,w)\ge  |T|/7$, this means that the integral contribution is less than $7$ times the fraction of  fractional contribution.\\

\noindent \textbf{Case 2:} When a cluster $\{u\}\cup T$ is returned. In this case, there are two sets of mistakes.
\begin{itemize}
    \item \textbf{The negative edges within the cluster.} In this case, the contribution of negative edge $(v,w)$, $v,w\in T$ to fractional disagreements is $\max\{x(v),x(w)\}-d(v,w) \ge x(v)-d(v,u)-d(u,w)\ge x(v) - 2\cdot \frac{2}{7}\ge 5/7-4/7 = 1/7$.

    \item \textbf{The positive edges to nodes outside the cluster.}
     For the positive edges, lets consider a node outside the cluster, $z\notin T\cup\{u\}$.
    If $d(u,z)\ge 3/7$, then $d(v,z) \ge d(u,z)-d(u,v) \ge 3/7- 2/7 = 1/7$
    
    If $2/7 <d(u,z)\leq 3/7$, we do the following:
    The total contribution of $z$ towards the integral component of the cluster objective is $|P|$  where P is the set of positive edges between the nodes of $T\cup\{u\}$ with $z$ and the number of negative edges incident on $z$ is $|Q| = |T|+1-|P|$. The fractional contribution of the edges incident on $z$ is 
    \begin{align*}
        &\sum_{w\in P}d(w,z)+\sum_{w\in Q}(\max\{x(w),x(z)\}-d(w,z))\\
        &\geq  \sum_{w\in P}(d(u,z)-d(u,w))+\sum_{w\in Q}(x(w)-d(u,w)-d(u,z))\\
       & \geq d(u,z)|P| + \sum_{w\in Q} (x(w) - d(u,z)) - \sum_{w\in P\cup Q}(d(u,w))\\
       &\geq d(u,z)|P| + |Q| (5/7 - d(u,z)) - \sum_{w\in \{u\}\cup T}(d(u,w))\\
       &\geq  d(u,z)|P| + |Q| (5/7 - d(u,z)) - \frac{|P|+|Q|}{7}\\
    \end{align*}
    This equation is a linear function in $d(u,z)$. So, we evaluate its values at the end points of the line to identify min and max.
    \begin{itemize}
        \item When $d(u,z)=2/7$, it evaluates to $|P|/7 + \frac{2}{7}|Q| > |P|/7$
        \item When $d(u,z)=3/7$, it evaluates to $\left(\frac{2}{7}\right)|P| + \frac{1}{7}|Q| > \frac{1}{7}|P|$
    \end{itemize}
\end{itemize}
This shows that the total integral disagreements of positive edges with any node $z\notin T$ is less than $7$ times the fractional disagreements of corresponding edges. Hence, the  approximation ratio of Algorithm \ref{alg:greedy} is $7$.

\subsection{Covering and Partitioning}
We use the same covering algorithm that uses the multiplicative weights algorithm from \cite{bansal2014min} along with the partitioning strategy to generate non-overlapping clusters. For completeness, we present the modified theorem statements for the complete graphs case.

\begin{thm}
Given a complete graph, running Algorithm \ref{Covering-Alg}\footnote{For complete graphs, the multiset $S^t$ in line 4 is generated using Algorithm \ref{alg:guess}} on the instance outputs a multiset $\mathcal{S}$ that satisfies the following conditions:
\begin{itemize}
\item $\forall S\in \mathcal{S}$
\begin{align*}
cost(S)\le 7 \cdot OPT
\end{align*}
\item $\forall v \in V$,
\begin{align*}
\frac{|\{S\in \mathcal{S}: v\in S\}|}{|\mathcal{S}|}\ge \frac{1}{5n k}
\end{align*}
\end{itemize}\label{thm:cover}
\end{thm}
\begin{proof}
Same as Proof of Theorem \ref{thm-covering}
\end{proof}



The covering generated by Algorithm \ref{Covering-Alg} is converted into a partitioning using Algorithm \ref{aggregation-alg}. The following result from  Section \ref{covering-aggregation-section} is used to bound to approximation ratio of the  generated partions. 

\begin{thm}
Given a complete graph  and a cover $\mathcal{S}$ consisting of subsets of $V$ such that:
\begin{itemize}
\item $\forall v\in V$, v is covered by at least $\frac{c}{nk}$ sets $S\in \mathcal{S}$ where $k$ is the number of partitions in the optimum solution which we guessed in the covering section and $c\in (0,1]$ and $cost(S)\leq B$
\end{itemize}
We propose a randomized algorithm which outputs a partition $\mathcal{P}$ of $V$ such that $\forall P_i\in \mathcal{P}, cost(P_i)\leq 2B$.\label{thm:split}
\end{thm}
\begin{proof}
Same as Theorem \ref{aggregation} for the positive edges. We can ignore the negative edges in this analysis as the cost of negative edges can never increase on splitting a cluster.
\end{proof}

Using Theorem  \ref{thm:cover} and \ref{thm:split}, we can generate a 14 approximation of the local correlation clustering problem for complete graphs.

\end{proof}

\section{Min-Max Constrained Multicut}
\label{appendix-constrained-mc}
In the following, first we prove that min-max constrained multicut is NP-Complete, and then we give a proof for Theorem \ref{min-max-constrained-MC}.
\subsection{NP-completeness of Min-Max Constrained Multicut}
\label{np-hard-proof-constrained-MC}
\begin{proof}
To show that min-max constrained multicut is NP-hard, we use a reduction from the min-max multiway cut problem introduced by Svitkina and Tardos~\cite{svitkina2004min}. In order to prove NP-hardness, we use the following reformulations of the min-max constrained multicut and min-max multiway cut problems.

\textbf{$D$-Multiway Cut}

Input: An undirected graph $G=(V,E)$ with edge capacities $c(e)\geq 0$ , and a set of nodes $X = \{x_1,\cdots,x_k\}$ called terminals, and an integer $D$.

Question: Is there a partitioning of the graph into $k$ parts, which separates all the terminals and the capacity of each part is at most $D$?

\textbf{$D$-Constrained Multicut}

Input: An undirected edge-weighted graph $G=(V,E)$, and a set of source-sink pairs $S_G = \{(s_1,t_1),\cdots, (s_T, t_T)\}$, an integer $k$ which is the minimum number of parts needed to separate all source-sink pairs, and an integer $D$.

Question: Is there a partitioning of the graph into $k$ parts, which separates all the source-sink pairs, and the capacity of each part(sum of weights of all edges with exactly one endpoint in that part) is at most $D$?

Given an instance $(G,X)$ of min-max multiway cut, we build a corresponding instance $(G',S_{G'}, k)$ of the min-max constrained multicut where $G' = G$ and for each pair of terminals $(x_i,x_j)$, we add a source-sink pair $(x_i,x_j)$ to $S_{G'}$, and put $k=|X|$. It could be seen that $G'$ has a $D$-constrained multicut if and only if $G$ has a $D$-multiway cut, and the NP-completeness proof is complete.

\end{proof}
\subsection{Approximation Algorithm}
Before explaining our result for min-max constrained multicut problem, we mention the following definition.
\begin{definition}
Demand of a vertex $v$ is the number of source sink pairs containing $v$.
\end{definition}
\begin{restatable}{thm}{minMaxConstrainedMC}
\label{min-max-constrained-MC}
Given an edge weighted graph $G=(V,E)$ on $n$ vertices, and a set of source sink pairs $S_G = \{(s_1,t_1),\cdots,(s_T,t_T)\}$, and a number $k$ which is the minimum number of parts needed to separate all source sink pairs, there exists a polynomial time algorithm which outputs a partitioning $\mathcal{P} = \{P_1,\cdots, P_{|\mathcal{P}|}\}$ of $G$ that separates all source sink pairs, and there is at least one terminal on each part. Additionally, $\max_{1\leq i \leq |\mathcal{P}|}\delta(P_i) \leq \mathcal{O}(\sqrt{\min\{T,n\}\cdot\Delta\cdot \log(n)\cdot \max\{\log(T),\log(k)\}})\cdot OPT$; where $OPT$ is the value of the optimum solution of min-max constrained multicut. $\Delta$ is the maximum demand of all the vertices.
\end{restatable}
We get improved approximation ratio for min-max constrained multicut on graphs excluding a fixed minor.
\begin{restatable}{thm}{Minor-MinMaxConstrainedMC}
\label{Minor-Constrained-MC}
Given an edge weighted graph $G$ excluding $K_{r,r}$ minors, there exists a polynomial time $\mathcal{O}(\sqrt{\min(T,n)\cdot\Delta}\cdot r^2)$-approximation algorithm for min-max constrained multicut.
\end{restatable}
In order to prove Theorem \ref{min-max-constrained-MC}, we use an approach similar to the one for min-max multicut with some modifications in the rounding and aggregation steps. Here we use the same SDP relaxation that we used for the min-max multicut. 

The SDP rounding procedure is consisting of two phases, the first phase is similar to what we did for the min-max multicut problem. After the last iteration of this phase, if $\eta(U)>H$ we output $F=S$ where $S$ is computed in the last iteration. In this case, the algorithm  terminates and we will not go through the next step.
If $\eta(U)\leq H$, the next step is a combination phase. Assume at the beginning of this phase, $U = \{S_1,S_2,\cdots,S_{\ell}\}$. While there are two sets $S_i$ and $S_j$ in $U$ such that $i\neq j$ and $vio(S_i\cup S_j) = vio(S_i) + vio(S_j)=0$, we make $S_i = S_i\cup S_j$ and remove $S_j$ from $U$. This means if there are no $(s,t)$ pairs such that $s\in S_i$ and $t\in S_j$, then $S_i$ and $S_j$ are combined.
After the combination phase is done, set $F = U$. In the proof of Theorem \ref{thm-find-one-set-multicut}, it is shown that for each sampled orthogonal separator, $\delta(S)\leq \frac{32D\cdot SDP\cdot \eta(S)}{H}$, where $D=\mathcal{O}(\sqrt{\log(n)\cdot\max\{\log(T),\log(k)\}})\cdot OPT$. It was also proved that $\eta(F)\leq 12H$. Therefore when we are combining multiple sampled orthogonal separators in $F$, the cut capacity of the combined set is at most $\mathcal{O}(D.SDP)$.

In Lemma \ref{bound-number-combined-orthogonal-separators}, it is proved at the end of the combination phase, $|U|\leq \sqrt{\min(2T,n)\cdot\Delta}$, and $\forall S_i \in U$, $vio(S_i) = 0$.

The covering procedure is same as the covering for min-max multicut, and at the end it can be proved that each vertex is covered by at least $\Omega(\frac{1}{k\sqrt{\min(T,n)\cdot\Delta}})$ fraction of the sets.

For aggregation, first we do Steps 1,2 in Algorithm \ref{aggregation-alg}. 
Next, we show how to combine non-terminal parts with terminal parts such that the number of cut edges on each part is at most: $\mathcal{O}(\sqrt{\min\{T,n\}\cdot\Delta\cdot \log(n)\cdot \max\{\log(T),\log(k)\}})\cdot OPT$.
\begin{lem}
\label{bound-number-combined-orthogonal-separators}
After the combination phase is done, $|U|\leq \sqrt{\min(2T,n)\cdot\Delta}$ .
\end{lem}
\begin{proof}
If $|U|=1$ the lemma is proved. Consider the case that $|U| > 1$. At the end of combination phase, each $S_i\in U$ has at least one terminal; otherwise it could have been combined with some other set in $U$. Assume $|U| > \sqrt{\min(2T,n)\cdot\Delta}$. Since the number of terminals is at most $\min(2T,n)$, there exists at least one part with less than $\frac{\min(2T,n)}{\sqrt{\min(2T,n)\cdot\Delta}}=\frac{\sqrt{\min(2T,n)\cdot\Delta}}{\Delta}$ terminals. Let's call this part $S_i$. Part $S_i$ could not have been merged with at most $(\frac{\sqrt{\min(2T,n)\cdot\Delta}}{\Delta}-1)\cdot \Delta$ other parts. Therefore, the total number of parts is at most:
 \[(\frac{\sqrt{\min(2T,n)\cdot\Delta}}{\Delta}-1)\cdot \Delta +1 = \sqrt{\min(2T,n)\cdot\Delta} - \Delta +1 \leq \sqrt{\min(2T,n)\cdot\Delta}\]
This is a contradiction since we assumed $|U| > \sqrt{\min(2T,n)\cdot\Delta}$.
\end{proof}

\begin{lem}
\label{low-violations-constrained-MC}
After the combination phase, for each $S_i \in U, vio(S_i)=0$.
\end{lem}
\begin{proof}
Consider $S_i\in U$ which is  a single orthogonal separator or is a combination of multiple orthogonal separators. Assume $S_i = S_1\cup S_2\cup\cdots\cup S_{\ell'}$. Therefore:
\[vio(S_i) = \sum_{j=1}^{\ell'} vio(S_j)=0\]
\end{proof}

\subsection{Covering \& Aggregation}
The algorithm for covering is the same as Algorithm \ref{Covering-Alg}. The following theorem can be proved and its proof is similar to the proof of Theorem \ref{thm-covering}.
\begin{thm}
\label{thm-covering-constrained-MC}
Given graph $G=(V,E)$ and $P(G)$ pairs of source and sink, running Algorithm \ref{Covering-Alg} on this instance outputs a multiset $\mathcal{S}$ that satisfies the following conditions:
\begin{itemize}
\item for all $S\in \mathcal{S}$: $\delta(S)\leq D\cdot OPT, \text{ where } D= \mathcal{O}(\sqrt{\log n\cdot\max\{\log(T),\log(k)\}}), vio(S) = 0.$
\item for all $v\in V$, $\frac{|\{S\in \mathcal{S}: v\in S\}|}{|\mathcal{S}|}\geq \frac{1}{17 k\sqrt{\min(T,n)\cdot\Delta}}$, where $k$ is the number of partitions of the optimum solution.
\end{itemize}
\end{thm}

\begin{thm}
\label{aggregation-constrained-MC}
Given a weighted graph $G=(V,E)$, a set of source-sink pairs $(s_1,t_1),\cdots,(s_T,t_T)$ and a cover $\mathcal{S}$ consisting of subsets of $V$ such that:
\begin{itemize}
\item $\forall v\in V$, $v$ is covered by at least a fraction $\frac{c}{k\sqrt{\min(T,n)\cdot\Delta}}$ of sets $S\in\mathcal{S}$, 
and $c=1/17$.
\item $\forall S\in\mathcal{S}$, $\delta(S)\leq B$, $vio(S)=0$.
\end{itemize}
We propose a randomized polynomial time algorithm which outputs a partition $\mathcal{P}$ of $V$ such that:
\begin{itemize}
\item $\forall P_i \in \mathcal{P}$, $\delta(P_i)\leq \mathcal{O}(\sqrt{\min(T,n)\cdot\Delta}B)$.
\item $vio(P_i)=0$.
\end{itemize}
\end{thm}
\begin{algorithm}[H]
\label{aggregation-alg-constrained-MC}
\SetAlgoLined
Step 1: Sort sets in $\mathcal{S}$ in a random order: $S_1,S_2,\cdots,S_{|\mathcal{S}|}$. Let $P_i = S_i\setminus (\cup_{j<i} S_j)$. \\
Step 2:
\While{There is a set $P_i$ such that $\delta(P_i)>2B$}{
	Set $P_i = S_i$ and for all $j\neq i$, set $P_j = P_j\setminus S_i$\;
}
Step 3: Let $B' = \max\{\frac{1}{k\sqrt{\min(2T,n)\cdot\Delta}}\sum_i \delta(P_i),2B\}$\\
\While{there are $P_i\neq \emptyset$, $P_j\neq \emptyset$ $(i\neq j)$ such that $\delta(P_i)+\delta(P_j)\leq B'$ and $P_i, P_j$ are non-terminal parts}{
	Set $P_i = P_i\cup P_j$ and $P_j = \emptyset$.
}
Step 4: For each terminal-part $P_i$, combine at most $2\sqrt{\min(2T,n)\cdot\Delta}$ non-terminal parts with it.
\caption{Aggregation Procedure For Min-Max Constrained Multicut}
\end{algorithm}
\begin{proof}
\label{aggregation-constrained-MC-proof-appendix}
A similar proof to the proof of Theorem \ref{aggregation} shows that Step 1, $\E[\sum_i \delta(P_i)]\leq 2k\sqrt{\min(2T,n)\cdot\Delta}B/c$. Along the same lines of proof of Theorem \ref{aggregation}, it can be shown the number of iterations in Step 2 is polynomially bounded and after the last iteration, $\E[\sum_i \delta(P_i)]\leq 2k\sqrt{\min(2T,n)\cdot\Delta}B/c$. 

After Step 3 is finished, each part $P_i$ satisfies $\delta(P_i)\leq B' = \max\{2B/c, 2B\} = 2B/c$, and $\sum_i \delta(P_i)\leq k\sqrt{\min(2T,n)\cdot\Delta}B'$, and if we merge any two sets $P_i$ and $P_j$, $\delta(P_i)+\delta(P_j)>B'$. Therefore, the total number of non-empty and non-terminal sets is at most $\frac{2k\sqrt{\min(2T,n)\cdot\Delta}B'}{B'} = 2k\sqrt{\min(2T,n)\cdot\Delta }$.

Next, we show the total number of parts with at least one terminal, which we call terminal parts, is at least $k$. If the total number of terminal parts is less than $k$, then there will be some violations in some of the terminal parts. However, since each part $P_i$ is a subset of a set $S$ such that $vio(S)=0$, therefore $vio(P_i) = 0$, which is a contradiction. 

Therefore, by combining an almost equal number of non-terminal parts with each terminal part, we can show for each part $P_i$, 
\[\delta(P_i)\leq \mathcal{O}(\sqrt{\min(T,n)\cdot\Delta}B') = \mathcal{O}(\sqrt{\min(T,n)\cdot\Delta}B) \]
\end{proof}

In order to prove Theorem \ref{Minor-Constrained-MC}, we just need to use LP separators instead of orthogonal separators and follow a similar approach that we used for proving Theorem \ref{MinorCC}.

\noindent
{\bf Acknowledgements:} We are grateful to Nikhil Bansal for
useful discussions during a Dagstuhl workshop on scheduling (18101) .
\bibliographystyle{abbrv}
\bibliography{bibfile}

\end{document}